\documentclass[aps,prb,preprint,groupedaddress,showpacs]{revtex4}
\usepackage{graphicx}
\begin{document}

\title{Size-dependent melting of spherical copper nanoparticles}

\author{Oleg A. Yeshchenko}
\author{Igor M. Dmytruk}
\affiliation{Physics Department, National Taras Shevchenko Kyiv University,
2/1 Akademik  Glushkov prosp., 03127 Kyiv, Ukraine}
\author{Alexandr A. Alexeenko}
\affiliation{Gomel State Technical University, Gomel, Belarus}
\author{Andriy M. Dmytruk}
\affiliation{Center for Interdisciplinary Research, Tohoku University,
 Aoba-ku, Aramaki Aza Aoba, 980-8578 Sendai, Japan}

\date{\today}

\begin{abstract}
We report size-dependent melting of spherical copper nanoparticles embedded into silica matrix. Based on the temperature dependence of the surface plasmon resonance energy and its width we observe two distinct melting regimes. For particles smaller than 20 nm the absorption spectrum changes monotonically with the temperature, and this allows us to assume the gradual solid-liquid phase transition (melting) of the nanoparticles or existence of superheated solid nanoparticles. In contrast, for nanoparticles larger than 20 nm, we observe a
jump-like increase of the bandwidth and non-monotonic dependence of surface plasmon energy at the temperatures below the bulk melting point. This indicates that the melting of large nanoparticles is a first-order phase transition similar to the melting of bulk copper.
\end{abstract}


\maketitle

\section{Introduction}
\label{intr}

Optical properties of copper nanoparticles are quite remarkable because the energy of the dipolar mode of surface collective electron plasma oscillations (surface plasmon resonance or SPR) coincides with the onset of interband transition. Therefore, optical spectroscopy gives an opportunity to study the particle-size dependence of both valence and conduction electrons. The intrinsic size effect in metal nanoparticles, caused by size and interface damping of the SPR \cite{1,Pinch}, is revealed experimentally by two prominent effects: a red shift of the surface plasmon band and its broadening. Scattering of the conduction electrons by the surface boundaries of a nanoparticle causes a dephasing of collective excitations, i.e. SPR, and is well described by the expression \cite{1}
\begin{equation}
\label{eq1}
\Gamma = \Gamma_{\infty} + A \frac {v_{\infty}}{r} ,
\end{equation}
where $\Gamma$ is width of plasmon absorption band, $\Gamma _{\infty} = 1/\tau _{\infty}$ is the size-independent damping constant which is related to the lifetimes of electron-electron, electron-phonon, and electron-defect scattering, $v_{\infty}$ is the Fermi velocity in bulk material ( $v_{\infty} = 1.57\times 10^{8}$ cm/sec in bulk copper), $r$ is the radius of nanoparticle, and $A$ is a theory-dependent parameter that includes details of the scattering process (e.g. isotropic or diffuse scattering \cite{1,Pinch,2,3}). Note, that Eq.(\ref{eq1}) describes the width of SPR band only when the onset of interband transitions has a higher energy than that of SPR energy, e.g. for silver nanoparticles. The confinement of conduction electrons in a small volume of a nanoparticle causes a blue shift of SPR band when the size of the nanoparticle is decreased, accompanied by a spill out effect beyond the physical boundary of the nanoparticle\cite{1}. The particle-size dependence of the interband transition edge is caused mostly by the surface stress-induced changes in the lattice constant and corresponding changes in the electronic band structure \cite{1}. This leads to an additional blue shift of SPR band with decreasing particle size.

Melting of crystalline materials is characterized by lost of the crystal order and following by a transformation to a liquid phase. As it has recently been shown, melting of small nanometer-size particles is quite different process than that of bulk materials (see, e.g. Refs. \onlinecite{4,ChPhL,5,6}). As the size of a particle decreases, the surface to volume ratio is increased inversely proportional to the radius of the particle. As a result, the surface atoms become more prone to detach from their positions and diffuse on the surface. Surface diffusion is followed by a partial melting of the nanoparticle mainly within a thin layer close to the surface, which is called surface melting or premelting \cite{7}. The surface melting generally occurs at temperatures much lower than the bulk melting but the details depend on the parameters such as size \cite{4}, morphology \cite{8},  defects \cite{9}, strain \cite{5}, and crystallographic orientation of the surface \cite{4,5}. Finally, similarly to the surface melting, the interior or volume melting of a nanoparticle takes place at melting temperature ($T_{m}$) lower than the bulk melting point. The decrease of $T_{m}$ with decreasing particle size is a well known phenomenon that has been described in terms of phenomenological thermodynamic models \cite{10}, and, more recently, simulated using molecular dynamics \cite{7,11,12,13}. Experimental evidence of the decrease of $T_{m}$ has been demonstrated by a variety of techniques \cite{ChPhL,14,15,16}.

We have shown earlier \cite{17} that absorption spectra of Cu/SiO$_{2}$ nanocomposite reveal a clear SPR  peak located around 560.8 nm (2.210 eV) - 581.1 nm (2.133 eV) depending on the size of Cu nanoparticles. The SPR peak position is in good agreement with that of predicted by the Mie theory. As we reported in the previous paper \cite{17}, the dependence of the width of plasmon peak on the nanoparticle diameter is perfectly described by the well known $1/r$ law (expression (\ref{eq1})). The fitting of our experimental dependence $\Gamma (d)$ obtained at the temperature of 293 K gives the following values of parameters (\ref{eq1}): $\Gamma _{\infty ,293} = 0.087$ eV and $A_{293} = 0.107$. 

Here we report the size dependent melting transition in spherical copper nanoparticles in a broad size range 5-65 nm by means of the absorption spectroscopy, which is, to our best knowledge, has not been reported so far. We noted already that the melting temperature of metals decreases when the size of the nanoparticles decreases. Since metal nanoparticles have been widely used in electronic and optoelectronic industry, the determination of the maximum allowed temperature for their exploitation is apparently a very important technological problem. This information is obtained from experiments of metal nanoparticles melting. Moreover, the optical spectroscopy is a nondestructive tool for determination of the extremal characteristics of exploitation regimes of nanometal composites based electronic devices. Besides the technological interest, the melting temperature is of fundamental scientific interest. In most reports the melting of nanoparticles has been detected by high resolution electron microscopy (HRTEM) and electron diffraction techniques. However, both techniques use the samples which are sufficiently heated by the electron beam. Therefore, there is no possibility to control the real temperature of the sample. On the other hand, the absorption spectroscopy gives such a possibility, i.e. the temperature determined in our work is the real temperatures of the sample. The study of the temperature dependences of the SPR energy and bandwidth in copper nanoparticles in broad size range have allowed us to reveal the two principally different regimes of melting of small and large copper nanoparticles.

\section{Fabrication of copper nanoparticles. Experimental procedures}
\label{exper}

Cu nanoparticles in a SiO$_{2}$ matrix were synthesized using a procedure described in Ref. \onlinecite{Alexeenko}. Porous silica matrices (obtained by a transformation to monolithic glasses) were produced by the conventional sol-gel technique based on hydrolysis of tetraethoxysilane (TEOS). We modified the procedure by introducing a doping followed by a chemical transformation of the dopants in air or controlled gaseous medium. A precursor sol was prepared by mixing of TEOS, water and alcohol, with the acid catalysts HNO3 or HCl. Silica powder with particle size about 5-15 nm (the specific surface area is 380 $\pm$ 30 m$^{2}$/g) was added into the sol followed by ultrasonication in order to prevent a large volume contraction during drying. The next gelation step resulted in formation of gels of desired shape. Gels were dried at room temperature under humidity control, and porous materials (xerogels) were obtained. Annealing in air at the temperature of 600 $^{o}$C during 1 h allowed us to control the porosity of SiO$_{2}$ matrices. The heating process resulted in visible variations of the density and specific surface area. Copper doping was performed by impregnation of xerogels in Cu(NO$_{3}$)$_{2}$ alcohol solution during 24 h. Then, the impregnated samples have been dried in air at 40 $^{o}$C during 24 h. Further processing of the Cu-doped xerogels was done in three different ways: (1) Annealing in air with gradual increase of the temperature from 20 $^{o}$C to 1200 $^{o}$C (annealing time at 1200 $^{o}$C: 5 min). (2) Initial annealing in air with gradual increase of the temperature from 20 $^{o}$C to 1200 $^{o}$C (annealing time at 1200 $^{o}$C: 5 min), then annealing in the atmosphere of molecular hydrogen at the temperature of 800 $^{o}$C during 1 h. (3) Annealing in H$_{2}$ with gradual increase of the temperature from 20 $^{o}$C to 1200 $^{o}$C (annealing time at 1200 $^{o}$C: 5 min). As we have shown in the previous work \cite{17}, the annealing in air leads to formation of both the copper nanoparticles and the copper oxide (Cu$_{2}$O) nanoparticles. An annealing in hydrogen results in reduction of Cu(I,II) to Cu(0) that is aggregated in the form of Cu nanoparticles \cite{17}. Glass samples were polished up to thickness about 1 mm for optical measurements. The samples obtained by annealing in air have low optical density, and are non-colored or slightly yellow (labeled hereafter as samples A); ones obtained at successive annealing in air and hydrogen have higher optical density, and are light-pink, orange or red colored (samples AH); ones obtained at annealing in hydrogen have the highest optical density, and have red color (samples H). 

A tungsten-halogen incandescent lamp was used as a light source for the absorption measurements. The single spectrometer MDR-3 has been used for \emph{in situ} measurements of absorption spectra. Absorption spectra of Cu/SiO$_{2}$ nanocomposites containing Cu nanoparticles in size range 5-65 nm have been measured in the temperature region of 77 - 910 K. The measurements at 77 K were performed for the samples dipped in the liquid nitrogen, and measurements at the room and higher temperatures were performed in air. The samples were placed into an open furnace during the absorption measurements at high temperatures. The thickness of the samples was controlled before and after heating and remained unchanged. The measured absorption spectra were unchanged after several cycles of heating and cooling, therefore, we conclude, that during the heating of samples performed at absorption measurements the copper particle diffusion in the matrix is absent. Therefore, the filling factor of nanoparticles in matrix and inter-particle distances did not change. So, particle diffusion can not be the cause of the effects observed in the temperature dependences of SPR bandwidth and energy considered in section \ref{res}. Transmission electron microscopy of studied Cu/SiO$_{2}$ nanocomposites was performed by JEOL JEM-2000EX electron microscope at a room temperature.

\section{Transmission electron microscopy of $\bf Cu/SiO_{2}$ nanocomposites}
\label{TEM}

\begin{figure}
\includegraphics{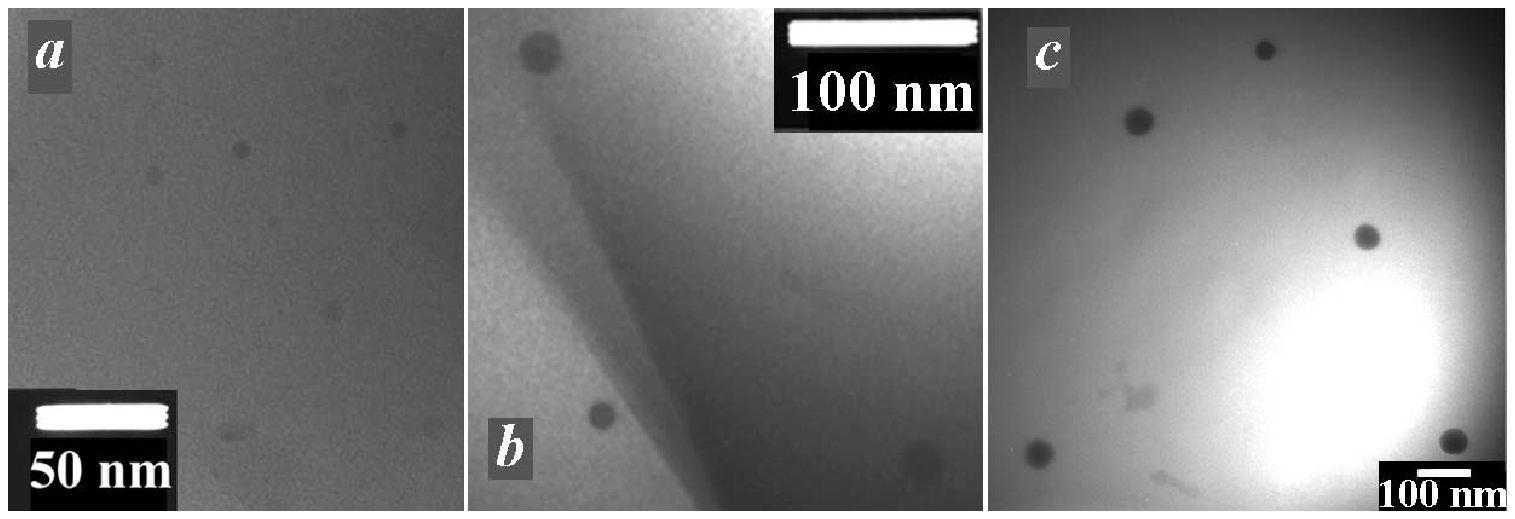}
\caption{\label{fig1} TEM images of Cu nanoparticles in silica matrix. $a$ - A1 sample annealed in air (mean diameter of copper nanoparticles $<d>$ = 6.1 nm), $b$ - AH2 sample annealed successively in air and hydrogen ($<d>$ = 16.7 nm), $c$ - H1 sample annealed in hydrogen ($<d>$ = 50.5 nm).}
 \end{figure}

Some TEM images of the nanoparticles synthesized at the different technology conditions are shown in fig. \ref{fig1}. The images prove the formation of spherical Cu nanoparticles of different sizes in the matrix depending on the conditions of annealing. The copper nanoparticles of small size (diameter $d$ is in the range of 2 - 15 nm) were grown in the samples annealed in air (samples A). The medium-sized nanoparticles ($d$= 15 - 40 nm) were grown in samples annealed in air and then in molecular hydrogen (samples AH). The largest particles ($d$= 40 - 65 nm) were grown in the samples annealed in the H$_{2}$ (samples H). Based on the analysis of TEM images we conclude that the distance between the nanoparticles is vastly larger than the size of the particles, therefore, the electromagnetic interaction between the particles is negligible \cite{1}. The particles have a spherical shape and are homogeneous - there is no evidence of a core-shell structure of the particles. Annealing in air leads to the oxidation of nanoparticles and formation of copper oxide (Cu$_{2}$O) \cite{17}. However, Cu$_{2}$O nanoparticles have not been observed in TEM images, most probably, due to their lower contrast as compared to the copper nanoparticles.

The analysis of TEM results shows that nanoparticles are characterized by Gaussian-like size distribution with clear-featured maximum. The dispersion of the distributions is quite small. Indeed, for sample A1 annealed in air $d$ = (6.1 $\pm$ 0.2) nm ($\varepsilon = \Delta d\, /<d> = $0.033), for sample AH1 annealed successively in the air and hydrogen $d$ = (33.9 $\pm$ 0.8) nm ($\varepsilon = \Delta d \, /<d> = $0.024), and for sample H$_{2}$ annealed in hydrogen $d$ = (51 $\pm$ 2) nm ($\varepsilon = \Delta d \, / <d> = $0.039). Because of the narrow size distribution of the nanoparticles its influence on the optical response of the nanoparticles can be neglected.

\section{Absorption of $\bf Cu/SiO_{2}$ nanocomposites. Results and discussion}
\label{res}

\begin{figure}
\includegraphics{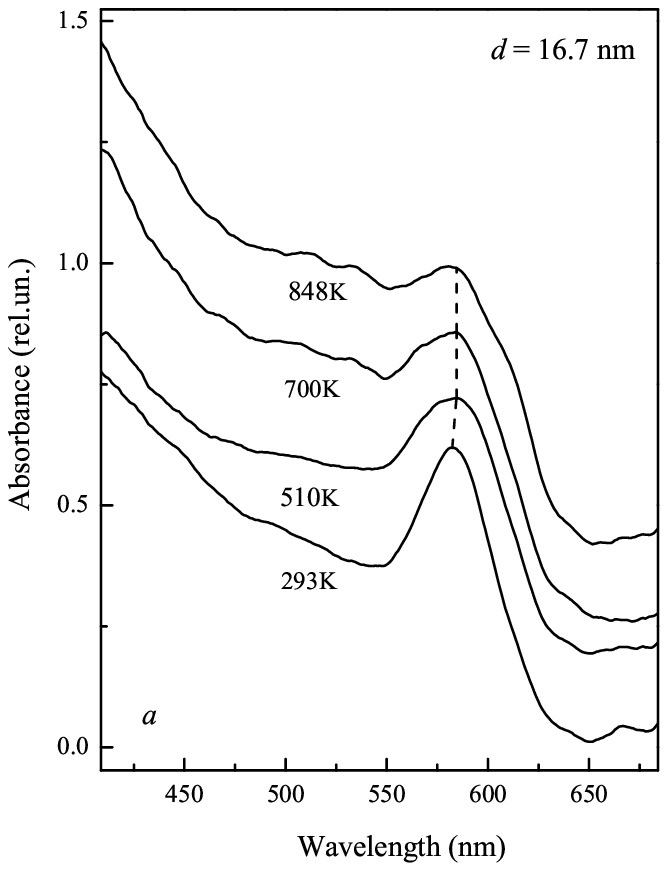}
\includegraphics{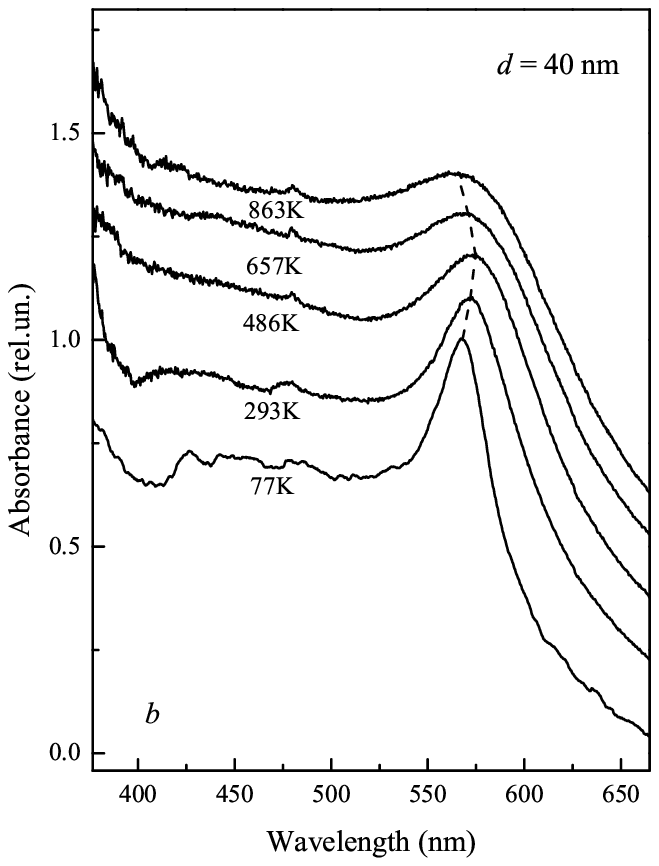}
\caption{\label{fig2} Absorption spectra of the copper nanoparticles with mean diameter of 16.7 nm (a) and 40 nm (b) in silica matrix measured at different temperatures.}
\end{figure}

The size-dependence of the width and energy of surface plasmon absorption band at the temperatures of 293 K and 77 K we reported earlier in \onlinecite{17}. In this paper we discuss the temperature dependence of SPR absorption bandwidth and its energy. The experimental absorption spectra of samples with average diameters of Cu nanoparticles 16.7 nm and 40 nm) at the different temperatures are shown in fig.\ref{fig2}. As one can see, the increase in the temperature leads to the broadening of the SPR peak. This broadening is well known for SPR in metal nanoparticles \cite{18}, in bulk metals \cite{19}, and in bulk semimetals (graphite) \cite{20}. The broadening is due to the increase of the frequency of the electron-phonon scattering ($\Gamma _{\infty}$) with increase of temperature. We have also observed a strong dependence of SPR energy on the temperature, see fig.\ref{fig2}.  

\begin{figure}
\includegraphics{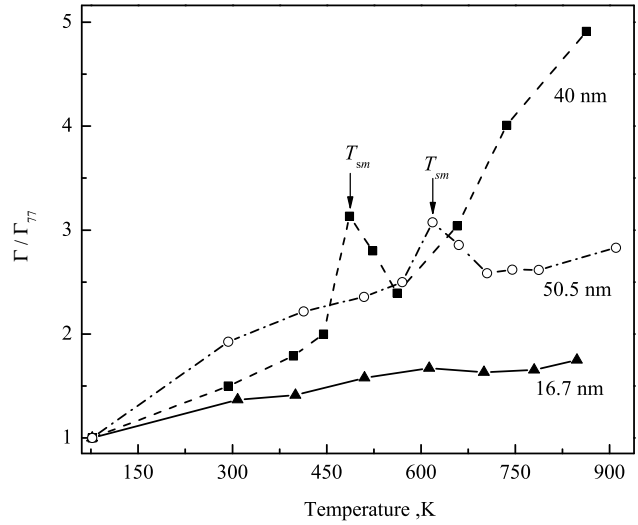}
\caption{\label{fig3} Temperature dependence of the surface plasmon absorption bandwidth (normalized to SP bandwidth at 77 K) for Cu nanoparticles of different sizes. Arrows mark the temperatures of the surface melting of nanoparticles.}
\end{figure}

Fig.\ref{fig3} depict the temperature dependences of SPR absorption bandwidth for small particles (with diameters smaller than 20 nm), where the gradual (without any features) increase of plasmon peak width with temperature takes place. This is similar to the dependences observed for SP in small (1.6 - 20 nm) gold nanoparticles \cite{18} and SP in bulk silver \cite{19}. The observed increase of SPR bandwidth for small Cu nanoparticles is close to linear dependence that has been predicted for the temperature dependence of frequency of electron-phonon scattering for temperatures higher than the Debye temperature \cite{21}: $\Gamma _{\infty} \propto T$. However, for larger (with the diameters larger than 20 nm) Cu nanoparticles we have observed a gradual increase of the bandwidth along with the abrupt significant increase (about 1.5 times) of the bandwidth with the increase of the temperature. Such an abrupt increase of the electron-phonon scattering frequency was reported in Ref. \onlinecite{22} at the transition of bulk metals from the solid to liquid phase, i.e. at the bulk metal melting. So, we can suppose that such jump-like increase of SPR bandwidth is caused by melting (surface melting as we show below) of the copper nanoparticles. Similar low-temperature melting of the copper nanorods (at the temperatures much lower than copper bulk melting point 1084.5 $^{o}$C) was observed in Ref. \onlinecite{23} by using TEM, SEM and XRD techniques. The ambient environment is an important factor affecting the process of nanoparticle melting. In our work we study the copper nanoparticles embedded in silica matrix. Therefore, it is reasonably to suppose that the wetting of the silica matrix by the molten copper should facilitate the melting that should decrease the melting temperature. It is well known that the corresponding abrupt changes must indicate a first-order phase transition. Let us emphasize that such feature in the temperature dependence of SPR bandwidth appears for rather large copper particles and is absent for small particles. We note that the temperature of the first-order transition in large copper nanoparticles decreases with decrease of the particle size, see fig.\ref{fig3}. 

\begin{figure}
\includegraphics{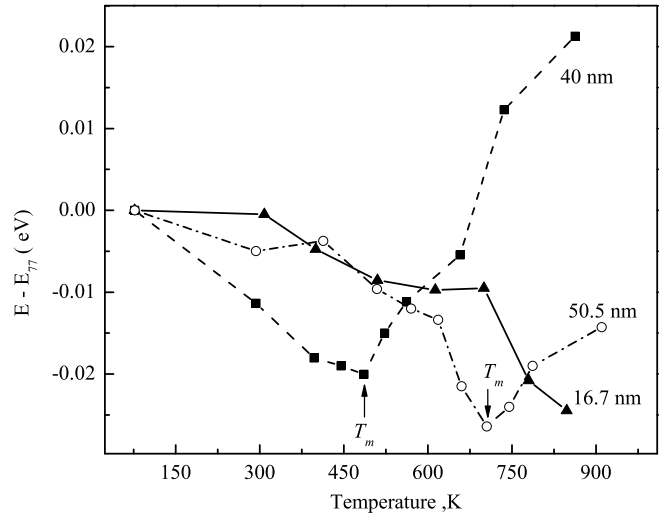}
\caption{\label{fig4} Temperature shift of the SP absorption band from the spectral position corresponding to 77 K for copper nanoparticles of different sizes. Arrows mark the temperatures of the melting of nanoparticles.}
\end{figure}

The measured temperature dependences of SPR energy (fig.\ref{fig4}) are different for the small and large Cu nanoparticles similar to the temperature dependence of plasmon bandwidth. For small Cu particles the monotonic red shift with the increase of temperature is observed, similar to small (1.6-20 nm) gold nanoparticles reported in Ref. \onlinecite{18}. For large copper nanoparticles the temperature dependence of SP absorption band energy demonstrates the non-monotonic character, namely at lower temperatures an absorption band is characterized by red shift with the increase of temperature and at higher ones the blue shift is observed. Let us note that the temperature (marked as $T_{m}$) at which the red shift changes to the blue one decreases with decrease of the particle size. 

\begin{figure}
\includegraphics{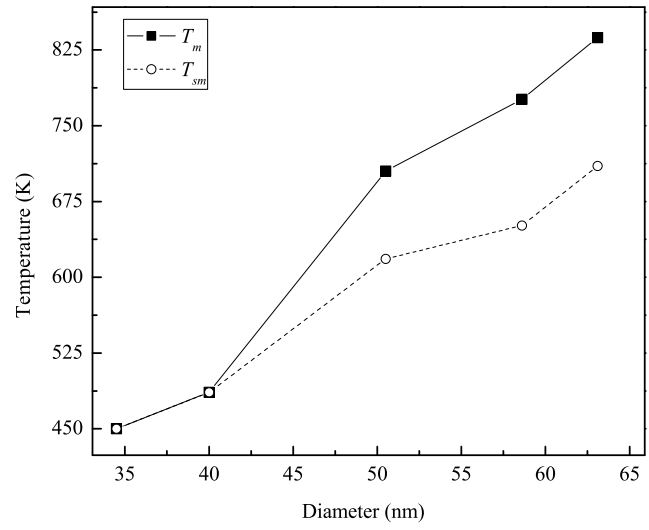}
\caption{\label{fig5} Dependences of the temperatures of melting and surface melting of copper nanoparticles on their diameter.}
\end{figure}

Let us discuss the temperature dependences of the SPR bandwidth and energy. The observed changes in spectral characteristics of SPR (energy and bandwidth) should manifest some transformation in the structure of nanoparticle. The transformation can be either the solid-liquid phase transition (melting) or the solid-solid phase transition (structural transition). The later was observed by electron diffraction technique for free copper nanoparticles \cite{Reinhard,Reinhard1} and by HRTEM technique for free gold nanoparticles \cite{Koga}. First, let us discuss the melting as possible reason for the observed temperature dependences of SPR spectral characteristics. As we mentioned earlier, the jump-like increase of the SPR bandwidth at certain temperatures suggests the first-order phase transition solid-liquid (melting) in copper nanoparticles with diameters larger than 20 nm. Above dependence is quite similar to the dependences observed at the melting of bulk metals \cite{22,24}. The melting of such kind begins by the process of the melting of thin layer close to the surface, i.e. surface melting. The surface melting generally occurs at temperatures much lower than the bulk melting point. A stable coexistence of spatially separated solid and liquid phase takes place during the increase of temperature up to the melting point. Melting of entire volume of the particle occurs at the volume melting temperature. The surface melting has to manifest itself in the SPR bandwidth mainly. The volume melting has to affect mainly on the energy of SPR as in copper the dipolar mode of SP resonance coincides with the dominant interband transition edge. One of the important characteristics of the bulk melted metals is the blue shift of the SP band occurring at the increase of the temperature \cite{24}. That effect is due to the reduction of polarizability of the melted ion core. In solid state the red shift of SPR energy is observed due to lattice dilation occurring at the increase of temperature. The lattice dilation decreases the electron density, and, hence, the plasma frequency $\omega _{p}= n_{e}e^{2}/\epsilon _{0}m_{e}$. Therefore, one can conclude that at the temperatures higher than temperature at which the red shift of SPR energy with the temperature changes to blue one the copper nanoparticles are melted, i.e. this temperature (marked as $T_{m}$) is the melting temperature of the Cu nanoparticle. Correspondingly, the temperature of jump-like increase of SP bandwidth is the temperature of surface melting (marked as $T_{sm}$). At the temperatures lower than $T_{sm}$ the nanoparticle is solid entirely. In the range $T_{sm} \leq T \leq T_{m}$ the coexistence of solid core and liquid surface occurs. The dependence of the temperatures of melting and surface melting on the diameter of Cu nanoparticle is shown in fig. \ref{fig5}. Let us note that these temperatures are equal for smaller particles with diameters up to 40 nm, and for larger particles the melting temperature exceeds the surface melting one. Most probably, that is due to greater role of surface in small particles. 

Let us discuss the solid-solid phase transition as the possible cause of the observed temperature dependences of SPR spectral characteristics. As it was shown in Ref. \onlinecite{Reinhard,Reinhard1} the morphology of small copper nanoparticles (smaller than about 4 nm) is icosahedral (Ih) or decahedral (Dh), where Ih and Dh are noncrystalline structures. The morphology of the particles larger than 4 nm can be Ih, Dh or fcc (face centered cubic), where fcc structure is crystalline structure, i.e. the structure of bulk copper. The similar situation occurs in gold nanoparticles \cite{Koga} where the morphologies of small nanoparticles (smaller than about 6 nm) are noncrystalline, namely icosahedral (Ih) or decahedral (Dh). The morphology of gold particles larger than 6nm can be Ih, Dh or fcc (face centered cubic), where fcc structure is crystalline structure of gold. The Ih and Dh structures are noncrystalline, i.e. these structures do not exist for bulk crystals. Since the copper nanoparticles can exist in form of three different structures (Ih, Dh and fcc), there is a possibility, that the observed peculiarities in temperature dependence of SPR bandwidth and energy at T$_{sm}$ and T$_{m}$ are the results of some solid-solid structure transition, i.e. Ih-Dh, Ih-fcc or Dh-fcc. However, the solid-solid transitions between noncrystalline Ih and Dh phases as well as transition from noncrystalline Ih or Dh phases to crystalline fcc one seems to be of low possibility due to the following reason; as it was shown in Refs.\onlinecite{Reinhard,Reinhard1,Koga} the relative fractions of the particles with noncrystalline Ih or Dh structures are vanishing for large particles with diameters higher than 20 nm, where the above mentioned peculiarities in temperature dependences of SPR energy and bandwidth are observed. For sizes higher than 20 nm nearly all the particles have the crystalline fcc structure. Therefore, we believe that the melting is a more probable reason for jump-like increase of SPR bandwidth at T$_{sm}$ and nonmonotonic behaviour of SPR energy at T$_{m}$ observed for large Cu nanoparticles than any kind of solid-solid structural transition. Since the temperature behavior of the energy and bandwidth of SPR at the melting of copper nanoparticles larger than 20 nm is the same as for bulk copper, one can assume that, similar to bulk copper, the large Cu particles have crystalline fcc morphology, which is in full agreement with the results of Refs.\onlinecite{Reinhard,Reinhard1,Koga}. Their melting occurs bulk-like, i.e. it is first-order phase transition. 

Thus, an abrupt jump-like increase of SPR bandwidth at $T_{sm}$ has to be attributed to the surface melting of studied large Cu nanoparticles. It is noted above that the temperature dependences of energy and bandwidth of SPR for small copper nanoparticles are quite different from ones for large particles. Namely, for small particles an increase of the bandwidth with temperature is gradual and  is not characterized by abrupt jump at certain temperature, and the temperature dependence of SPR energy is monotonically decreasing. Both these dependences are similar to the corresponding dependences observed e.g. for small gold nanoparticles \cite{18}. Let us note that temperatures reached in experiments with the small Cu nanoparticles are the same as temperatures at which evidences for large Cu nanoparticles melting have been obtained. So, as the melting temperature is lower for smaller particles, one can suppose that the melting of small nanoparticles occurs in other way than for large ones. The molecular dynamics simulations of the solid-liquid phase transitions in various metal nanoparticles were performed in Refs. \onlinecite{7,11,12,13}, where it was  shown that two characteristic particle sizes exist. For particles with size larger than first one ($d_{2}$) the solid-liquid phase coexistence is stable. The melting of such particles is first-order transition, i.e. the melting has to occur likely to bulk metals. For particles with size smaller than second characteristic value ($d_{1}$) the solid-liquid coexistence is unstable. It is shown in Ref. \onlinecite{12} that such small particles can be superheated and non-melted up to temperatures exceeding the bulk melting temperature. Finally, in the middle-sized particles the coexistence of the solid and liquid phases is metastable, and dynamic and transient coexistence can exist. Taking into account the results of Refs. \onlinecite{Reinhard,Reinhard1,Koga}, one can suppose the origin of the critical size of about 20 nm obtained for our copper nanoparticles. As it was shown in these studies, the relative fraction of both copper and gold nanoparticles with fcc crystalline morphology increases with the increase of the particle size, and on the contrary the relative fractions of the noncrystalline Ih and Dh morphologies decrease. In the region of size higher than 20 nm the nanoparticles with Ih and Dh structures are absent, and the nanoparticles with fcc structure only exist. Therefore, it is plausible to assume that, since the copper nanoparticles larger than about 20 nm have the crystalline fcc morphology, their melting would occur bulk-like. The copper particles smaller than about 20 nm have mainly the noncrystalline Ih or Dh morphologies. Therefore, their melting occurs in other way than in bulk-like large particles. So, one can suppose that the size of about 20 nm is the $d_{2}$ characteristic size mentioned above. 

Therefore, taking into account our experimental data, data of Ref. \onlinecite{Reinhard,Reinhard1,Koga} and data of computer simulations \cite{11,12} one can conclude that the melting of large studied Cu nanoparticles (with diameters larger than 20 nm) is first-order transition, and correspondingly, it is similar to the melting of bulk copper, i.e. the stable solid-liquid phase coexistence occurs in the temperature range from surface melting to volume melting. And, most probably, Cu nanoparticles with diameters up to 20 nm are characterized by the unstable or metastable solid-liquid phase coexistence. So, the observed monotonic temperature dependences of the surface plasmon bandwidth and energy suggest either the existence of the solid superheated particles or the fact that melting of such particles occurs gradually, and it is not first-order transition, that is quite different from the melting of bulk copper. We note, that the results of molecular dynamics that we use in our analysis of the observed phase changes are conducted for isolated nanoparticles whereas our Cu particles are embedded in a matrix. In principle, the behavior of isolated nanoparticles can differ to embedded particles. However, we think that environment (silica matrix) can change the values of melting characteristics (critical size of nanoparticle, melting temperature), but the qualitative behavior of the nanoparticle at melting should remain the same.

\section{Conclusions}
\label{concl}

In conclusion, we studied the temperature dependences of the bandwidth and energy of surface plasmon in spherical copper nanoparticles embedded in silica matrix. The obtained dependences indicate that for small particles (smaller than 20 nm) either the solid superheated nanoparticles exist or the melting is not a first-order phase transition that is different from the melting of bulk copper. The melting of large Cu particles (larger than 20 nm) is a first-order transition accompanying by jump-like increase of surface plasmon bandwidth and non-monotonic dependence of surface plasmon energy with the temperature that is similar to bulk copper melting.

\section{Acknowledgment}
\label{acknow}

The authors thank Dr. A.Pinchuk for helpful discussions and for careful proofreading of the manuscript.

\end{document}